\documentclass[aps,prb,a4paper,preprint,showpacs]{revtex4}
\usepackage{amsmath}
\usepackage{amsfonts}
\usepackage{amssymb}
\usepackage{slashbox}
\usepackage{graphicx}
\usepackage{amsbsy}

\begin{document}
\title{Pumped shot noise in adiabatically modulated
graphene-based double-barrier structures}
\author{Rui Zhu\renewcommand{\thefootnote}{*} \footnote{Corresponding author. Electronic address:
rzhu@scut.edu.cn} and Maoli Lai  }
\address{Department of Physics, South China University of Technology,
Guangzhou 510641, People's Republic of China }

\begin{abstract}

Quantum pumping processes are accompanied by considerable quantum
noise. We investigated the pumped shot noise (PSN) properties in
adiabatically modulated graphene-based double-barrier structures.
General expressions for adiabatically PSN in phase-coherent
mesoscopic conductors are derived based on the scattering approach.
It is found that comparing with the Poisson processes, the PSN is
dramatically enhanced where the dc pumped current changes flow
direction, which demonstrates the effect of the Klein paradox.

\end{abstract}


\maketitle

\narrowtext

\section{Introduction}

Quantum pumping is a transport mechanism which induces dc charge and
spin currents in a nano-scale conductor in the absence of a bias
voltage by means of a time-dependent control of some system
parameters. Research on quantum pumping has attracted continued
interest since its prototypical proposition due to its importance in
quantum dynamic theory and potential application in various
fields\cite{Ref40, Ref2, Ref3, Ref4, Ref5, Ref6, Ref7, Ref8, Ref9,
Ref10, Ref11, Ref12, Ref13, Ref14, Ref15, Ref16, Ref17, Ref18,
Ref19, Ref20, Ref21, Ref22, Ref23, Ref24, Ref25, Ref26, Ref27,
Ref28, Ref29, Ref31, Ref33, Ref36, Ref37, Ref41, Ref44, Ref45,
Ref46, Ref47}. The pumped current (PC) and noise properties in
various nano-scale structures were investigated such as the
magnetic-barrier-modulated two dimensional electron gas\cite{Ref5},
mesoscopic one-dimensional wire\cite{Ref7, Ref23}, quantum-dot
structures\cite{Ref6, Ref12, Ref13, Ref29, Ref30, Ref37}, mesoscopic
rings with Aharonov-Casher and Aharonov-Bohm effect\cite{Ref8},
magnetic tunnel junctions\cite{Ref11}, chains of tunnel-coupled
metallic islands\cite{Ref26}, the nanoscale helical
wire\cite{Ref27},the Tomonaga-Luttinger liquid\cite{Ref25}, and
garphene-based devices\cite{Ref21, Ref22, Ref41, Ref44, Ref45,
Ref46, Ref47}.

Graphene continues to attract intense interest, especially as an
electronic system in which charge carriers are Dirac-like particles
with linear dispersion and zero rest mass\cite{Ref42}. Quantum
pumping properties of graphene-based devises have been investigated
by several groups\cite{Ref21, Ref22, Ref41, Ref44, Ref45, Ref46,
Ref47}. It is found that the direction of the PC can be reversed
when a high potential barrier demonstrates stronger transparency
than a low one as an effect of the Klein paradox\cite{Ref21}. The
shot noise properties of a quantum pump are important in two
aspects: understanding the underlying mechanisms of the shot noise
may offer possible ways to improve pumping efficiency and achieve
optimal pumping. On the other hand, the shot noise reflects current
correlation and is sensitive to the pump source
configuration\cite{Ref43}. The pumped shot noise (PSN) properties
may provide further information of the correlation between the
transport Dirac Fermions of graphene governed by the Klein paradox
and electron chirality. However, this topic has not ever been looked
into. In this work, we focus on the PSN properties in adiabatically
modulated graphene-based double-barrier structures based on general
expressions we derived from the scattering approach. The effect of
the Klein paradox on the PSN is illuminated.

\section{Theoretical formulation}

The crystal structure of undoped graphene layers is that of a
honeycomb lattice of covalent-bond carbon atoms. One valence
electron corresponds to one carbon atom and the structure is
composed of two sublattices, labeled by A and B. In the vicinity of
the ${\bf{K}}$ point and in the presence of a potential $U$, the
low-energy excitations of the gated graphene monolayer are described
by the two-dimensional (2D) Dirac equation
\begin{equation}
 v_F \left( {{\mathbf{\sigma}} \cdot {\bf{\hat p}}} \right)\Psi  = \left( {E - U} \right)\Psi ,
\end{equation}
where the pseudospin matrix $\vec \sigma $ has components given by
Pauli's matrices and ${\bf{\hat p}} = (p_x ,p_y )$ is the momentum
operator. The ``speed of light" of the system is $v_{F}$, i.e., the
Fermi velocity ($v_F  \approx 10^6 $ m/s). The eigenstates of Eq.
(1) are two-component spinors $\Psi  = [\psi _A ,\psi _B ]^T $,
where $\psi _A $ and $\psi _B $ are the envelope functions
associated with the probability amplitudes at the respective
sublattice sites of the graphene sheet.

In the presence of a one-dimensional confining potential $U=U(x)$,
we attempt solutions of Eq. (1) in the form $\psi _A (x,y) = \phi _A
(x)e^{ik_y y} $ and $\psi _B (x,y) =i \phi _B (x)e^{ik_y y} $ due to
the translational invariance along the $y$ direction. The resulting
coupled, first-order differential equations read as
\begin{equation}
 d\phi _B /d\xi  + \beta \phi _B  = (\varepsilon  - u )\phi _A ,
\end{equation}
\begin{equation}
d\phi _A /d\xi  - \beta \phi _A  =  - (\varepsilon  - u )\phi _B .
\end{equation}
Here $\xi =x/L$, $\beta =k_y L$, $u = UL/\hbar v_F $, and
$\varepsilon =EL/\hbar v_F $ ($L$ is the width of the structure).
The incident angle $\theta $ is given by $\texttt{sin}(\theta)=
\beta / \varepsilon$. We consider a double-barrier structure with
two square potentials of height $U_1$ and $U_2$, which can be time
dependent modulated by ac gate voltages (see fig. 1). Eqs. (2) and
(3) admit solutions which describe electron states confined across
the well and propagating along it. As typical values $L/4$ for the
barrier widths and the inter-barrier separation $L/2$ are used, the
transmission and reflection amplitude $t$ and $s$ are determined by
matching $\phi _{A}$ and $\phi _{B}$ at region interfaces.

Following the standard scattering approach\cite{Ref3, Ref4} we
introduce the fermionic creation and annihilation operators for the
carrier scattering states. The operator $ \hat a_{L }^\dag (E,
\theta,t ) $ or $ \hat a_{L} (E, \theta ,t) $ creates or annihilates
particles with total energy $E$ and incident angle $\theta $ in the
left lead at time $t$, which are incident upon the sample.
Analogously, we define the creation $ \hat b_{L }^\dag (E, \theta
,t) $ and annihilation $ \hat b_{L } (E, \theta ,t) $ operators for
the outgoing single-particle states. Considering a particular
incident energy $E$ and incident angle $\theta$, the scattering
matrix $s$ follows from the relation
\begin{equation}
\left( {\begin{array}{*{20}c}
   {b_{L } }  \\
   {b_{R  } }  \\
\end{array}} \right) = \underbrace {\left( {\begin{array}{*{20}c}
   R & {T'}  \\
   T & {R'}  \\
\end{array}} \right)}_{\hat s}\left( {\begin{array}{*{20}c}
   {a_{L  } }  \\
   {a_{R  } }  \\
\end{array}} \right),
\end{equation}
where, $T$ and $R$ are the scattering elements of incidence from the
left reservoir and $T'$ and $R'$ are those from the right reservoir.

The frequency of the potential modulation is small compared to the
characteristic times for traversal and reflection of electrons and
the pump is thus adiabatic. In this case one can employ an instant
scattering matrix approach, i.e. ${\hat s} (t)$ depends only
parametrically on the time $t$. To realize a quantum pump one varies
simultaneously two system parameters, e.g. \cite{Ref3, Ref4}
\begin{equation}
\begin{array}{l}
 X_1 \left( t \right) = X_{10}  + X_{\omega ,1} e^{i\left( {\omega t - \varphi _1 } \right)}  + X_{\omega ,1} e^{ - i\left( {\omega t - \varphi _1 } \right)} , \\
 X_2 \left( t \right) = X_{20}  + X_{\omega ,2} e^{i\left( {\omega t - \varphi _2 } \right)}  + X_{\omega ,2} e^{ - i\left( {\omega t - \varphi _2 } \right)} . \\
 \end{array}
 \end{equation}
Here, $X_1$ and $X_2$ are measures for the two time-dependent
barrier heights $U_1$ and $U_2$ (see Fig. 1), which can be modulated
 by applying two low-frequency ($\omega$)
alternating gate voltages.  $X_{\omega ,1} $ and $X_{\omega ,2} $
are the corresponding oscillating amplitudes with phases
$\varphi_{1/2}$;
  $X_{10}$ and $X_{20}$ are the static (equilibrium) components. The scattering
  matrix $\hat s$
being a function of parameters $X_{j} (t)$ depends on time.

We suppose an adiabatic quantum pump, i.e., the external parameter
changes so slowly that up to corrections of order $\hbar \omega /
\gamma$ ( $\gamma$ measures the escape rate), we can apply an
instant scattering description using the scattering matrix $ {\hat
s} \left( t \right)$ frozen at some time $t$. Usually the varying of
the wave is sufficiently smooth on the scale of the dwell time. And
we assume that the amplitude ${X_{\omega ,j} }$ is small enough to
keep only the terms linear in ${X_{\omega ,j} }$ in an expansion of
the scattering matrix\cite{Ref4}
\begin{equation}
 \hat s\left( t
\right) \approx \hat s^0 + \hat s^{ - \omega } e^{i\omega t}  + \hat
s^{ + \omega } e^{ - i\omega t} .
\end{equation}
In the limit of small frequencies the amplitudes $\hat s^{ \pm
\omega } $ can be expressed in terms of parametric derivatives of
the on-shell scattering matrix $\hat s$,
\begin{equation}
\hat s^{ \pm \omega }  = \sum\limits_j {X_{\omega ,j} e^{ \pm
i\varphi _j } \frac{{\partial \hat s}}{{\partial X_j }}} .
\end{equation}
The expansion, Eq. (6), is equivalent to the nearest sideband
approximation which implies that a scattered electron can absorb or
emit only one energy quantum $\hbar \omega$ before it leaves the
scattering region.

The problem of current noise in a quantum pump is closely connected
with the problem of quantization of the charge pumped in one cycle.
On the other hand, the noise in mesoscopic phase-coherent conductors
is interesting in itself because it is very sensitive to quantum
mechanical interference effects and can give additional information
about the scattering matrix\cite{Ref4}. To describe the
current-current fluctuations we will use the correlation
function\cite{Ref48}
\begin{equation}
S_{\alpha \beta } \left( {t,t'} \right) = \frac{1}{2}\left\langle
{\Delta \hat I_\alpha  \left( t \right)\Delta \hat I_\beta  \left(
{t'} \right) + \Delta \hat I_\beta  \left( {t'} \right)\Delta \hat
I_\alpha  \left( t \right)} \right\rangle ,
\end{equation}
with $\Delta \hat I = \hat I - \left\langle {\hat I} \right\rangle $
and $\hat I_\alpha  \left( t \right)$ is the quantum-mechanical
current operator in the lead $\alpha $ as
\begin{equation}
\hat I_\alpha  \left( t \right) = \frac{e}{h}\left[ {\hat b_\alpha
^\dag  \left( t \right)\hat b_\alpha  \left( t \right) - \hat
a_\alpha ^\dag  \left( t \right)\hat a_\alpha  \left( t \right)}
\right].
\end{equation}
The time-dependent operator is $\hat a_\alpha  \left( t \right) =
\int {dE\hat a_\alpha  \left( E \right)e^{{{ - iEt} \mathord{\left/
 {\vphantom {{ - iEt} \hbar }} \right.
 \kern-\nulldelimiterspace} \hbar }} } $ and $\hat b_\alpha  \left( t \right) =
 \sum\limits_\beta  {s_{\alpha \beta } \left( t \right)\hat a_\beta  \left( t \right)} $
with ${s_{\alpha \beta } }$ an element of the instant scattering
matrix $\hat s$. Note that in the case of a time-dependent scatterer
the correlation function depends on two times $t$ and $t'$. Here we
are interested in the noise averaged over a long time ($\Delta t \gg
{{2\pi } \mathord{\left/
 {\vphantom {{2\pi } \omega }} \right.
 \kern-\nulldelimiterspace} \omega }$) and we investigate
\begin{equation}
 S_{\alpha \beta } \left( t \right) = \frac{\omega }{{2\pi }}\int_0^{{{2\pi } \mathord{\left/
 {\vphantom {{2\pi } \omega }} \right.
 \kern-\nulldelimiterspace} \omega }} {dtS_{\alpha \beta } \left( {t,t'}
 \right)}.
\end{equation}
In addition we restrict our consideration to the zero-frequency
component of the noise spectra $S_{\alpha \beta }  = \int
{dtS_{\alpha \beta } \left( t \right)} $. Substituting the current
operator Eq. (9), and taking into account Eqs. (4) and (6) we can
write the time-averaged zero-frequency PSN as
\begin{equation}
\begin{array}{c}
 S_{\alpha \beta }  = \frac{{e^2 \omega }}{{2\pi }}\sum\limits_{\mu \nu j_1 j_2 } {X_{\omega ,j_2 } X_{\omega ,j_1 } s_{\upsilon \beta }^{\dag 0} \frac{{\partial s_{\alpha \nu } }}{{\partial X_{j_1 } }}\frac{{\partial s_{\beta \mu } }}{{\partial X_{j_2 } }}s_{\mu \alpha }^{\dag 0} \cos \left( {\varphi _{j_1 }  - \varphi _{j_2 } } \right)}  \\
  \hspace{0.8cm} + \frac{{e^2 \omega }}{{2\pi }}\sum\limits_{\mu \nu j_1 j_2 } {X_{\omega ,j_2 } X_{\omega ,j_1 } s_{\upsilon \beta }^{\dag 0} s_{\alpha \nu }^0 \frac{{\partial s_{\beta \mu } }}{{\partial X_{j_2 } }}\frac{{\partial s_{\mu \alpha }^\dag  }}{{\partial X_{j_1 } }}\cos \left( {\varphi _{j_1 }  - \varphi _{j_2 } } \right)}  \\
  \hspace{0.8cm} + \frac{{e^2 \omega }}{{2\pi }}\sum\limits_{\mu \nu j_1 j_2 } {X_{\omega ,j_2 } X_{\omega ,j_1 } \frac{{\partial s_{\upsilon \beta }^\dag  }}{{\partial X_{j_2 } }}\frac{{\partial s_{\alpha \upsilon } }}{{\partial X_{j_1 } }}s_{\beta \mu }^0 s_{\mu \alpha }^{\dag 0} \cos \left( {\varphi _{j_1 }  - \varphi _{j_2 } } \right)}  \\
  \hspace{0.8cm} + \frac{{e^2 \omega }}{{2\pi }}\sum\limits_{\mu \nu j_1 j_2 } {X_{\omega ,j_2 } X_{\omega ,j_1 } \frac{{\partial s_{\upsilon \beta }^\dag  }}{{\partial X_{j_2 } }}s_{\alpha \upsilon }^0 s_{\beta \mu }^0 \frac{{\partial s_{\mu \alpha }^\dag  }}{{\partial X_{j_1 } }}\cos \left( {\varphi _{j_1 }  - \varphi _{j_2 } } \right)}  \\
  \hspace{-0.4cm} + \frac{{e^2 \omega }}{{2\pi }}\sum\limits_{\mu \upsilon j_1 j_2 j_3 j_4 } {\left[ {X_{\omega ,j_1 } X_{\omega ,j_4 } X_{\omega ,j_2 } X_{\omega ,j_3 } \frac{{\partial s_{\beta \mu } }}{{\partial X_{j_4 } }}\frac{{\partial s_{\mu \alpha }^\dag  }}{{\partial X_{j_1 } }}} \right.}  \\
 \hspace{0.8cm} \left. { \times \frac{{\partial s_{\alpha \upsilon } }}{{\partial X_{j_2 } }}\frac{{\partial s_{\upsilon \beta }^\dag  }}{{\partial X_{j_3 } }}\cos \left( {\varphi _{j_4 }  - \varphi _{j_1 }  + \varphi _{j_3 }  - \varphi _{j_2 } } \right)} \right]. \\
 \end{array}
\end{equation}
Eq. (11) is the central result of this manuscript, which can be used
to investigate the time-averaged zero-frequency PSN properties in
different nanoscale adiabatic pumping structures. Detailed
derivation is provided in the Appendix A.

The PC could be expressed in terms of the scattering matrix as
follows\cite{Ref4, Ref21}.
\begin{equation} I_\alpha   =
\frac{{e\omega }}{{2\pi }}\sum\limits_{\beta j_1 j_2 } {X_{\omega
,j_1 } X_{\omega ,j_2 } \frac{{\partial s_{\alpha \beta }
}}{{\partial X_{j_1 } }}\frac{{\partial s_{\alpha \beta }^*
}}{{\partial X_{j_2 } }}2i\sin \left( {\varphi _{j_1 }  - \varphi
_{j_2 } } \right)}.
\end{equation}
Due to current conservation, it can be seen that for a two-lead
(left and right) quantum pump (see Fig. 1), $I_{L}=I_{R}$ and
$S_{LL} = S_{RR} = - S_{LR} = - S_{RL} $. It is reasonable to
consider only the $I_{L}$ and $S_{LL}$. The symbols $I_{p}$ and
$S_{p}$ are used for the PC $I_{L}$ and PSN $S_{LL}$, respectively.
A convenient measure for the relative noise strength is the Fano
factor defined by $F_{p}=S_{p}/2eI_{p}$, which characterizes the
noise with respect to the Poisson processes. The Poissonian shot
noise in the configuration of a quantum pump is discussed in the
Appendix B.

\section{Numerical results and interpretations}

We consider the PSN properties in the graphene-based conductor
modulated by two ac gate voltages sketched in Fig. 1. In numerical
calculations, the parameters $U_{10}=U_{20} =100$ meV, $L=200$ nm,
$U_{1 \omega}=U_{2 \omega}=0.01$ meV. The phase difference of the
two oscillating gate potentials $\phi  = \varphi _2  - \varphi _1 $
in the radian unit.

The PC, PSN, and Fano factor as functions of the incident angle
$\theta $ for different Fermi energies are shown in Fig. 2.
Electrons at the Fermi levels of the reservoirs are driven to
flow in one direction by modulating the two barriers with a phase
lag, which results in a dc PC at zero bias. The
direction of the PC can be reversed when a high potential barrier
demonstrates stronger transparency than a low one, which results
from the Klein paradox\cite{Ref21}. The PSN is nonnegative as it
measures the PC-PC correlation flowing in the same direction. It can
be seen that the PSN increases when the PC is increased. The Poisson shot noise demonstrates the process governed by uncorrelated electrons and barrier gates without conduction structure (see the Apendix B). In graphene conductors, quantum states below potential barriers are hole states. Transmission from electron states outside the potential barriers into the hole states inside the potential barriers is characterized by the Klein paradox. For some incident angles and certain potential heights when chirality meets, the potential barrier is transparent. For other situations violating chirality alignment, the potential barrier is opaque. As the ac drivers modulate the potential barriers in time, the transmission is varied and a dc current is pumped from one reservoir to the other. Klein paradox virtually correlates the hole states with the electron states. Therefore, the PSN is remarkably enhanced beyond the Poisson value, the latter of which indicates uncorrelated transport. The PSN relative to the Poisson value measured by the Fano factor is presented in Fig. 2 (c). It can be seen that the Fano factor is above 1. Klein paradox induced virtual correlation between electrons and holes enhances the PSN beyond the Poisson value. It is also revealed in Fig. 2 that the PSN and Fano factor are extremely large at the incident angle when the PC reverses direction. At those incident angles, the chirality alignment is reversed, which induces extraordinary correlation between electrons and holes in virtual transport processes.

The PC, PSN, and Fano factor as functions of the Fermi energy of the two reservoirs
$E $ for the incident angle $\theta =0.01$ are shown in Fig. 3. The absolute value of the PC is in maximums at transmission peaks of the two-barrier graphene structure. Around the transmission peaks, the PC reverses direction. In our pumping configuration, ${\varphi _1} < {\varphi _2}$. The right gate opens in advance of the left gate. In quantum pumps constructed by other conductors, the PC always flows from the right to the left reservoir at the ${\varphi _1} < {\varphi _2}$ phase lag. As a result of the Klein paradox, higher potential barrier demonstrates stronger transmission when the chirality alignment meets and the PC reverses direction. The chirality consistency favoring transmission is different between the incident energy above and below the peak energy. When the Fermi energy is smaller than the Dirac point 100 meV, above the peak energy, higher potential barrier demonstrates stronger transmission and the PC flows from the left reservoir to the right. Below the peak energy, higher potential barrier demonstrates weaker transmission and the PC flows from the right reservoir to the left. When the Fermi energy is larger than the Dirac point, the PC direction is reversed as the transmission configuration is reversed. Larger PCs have relatively stronger current-current correlation. The shot noise demonstrates peaks at the PC peaks as shown in Fig. 3 (b). The shot noise is positive since the rightward current flow correlates with the rightward current flow and vice versa. The Fano factor is above 1 due to the Klein paradox induced virtual correlation between electrons and holes. At energies when the PC reverses direction, the shot noise is extraordinarily enhanced beyond the Poisson value. At those energies, the chirality alignment is reversed, which induces extraordinary correlation between electrons and holes in virtual transport processes.

The PC, PSN, and the Fano factor as
functions of the driving phase difference are shown in Fig. 4. The PC varies with the driving phase $\phi $ in sinusoidal function and the PSN in cosinusoidal function, which can be already seen in Eqs. (11) and (12). The last term of Eq. (11) is a product of four pumping amplitudes, four derivatives of the scattering-matrix elements relative to the oscillating parameter, and a $\cos 2\phi $ function. As small pumping amplitudes are considered in our approach, the magnitude of this term is negligible. Therefore, the PSN is a function of $\cos \phi $ and no $\cos 2\phi $-form modulation is observable. From Fig. 4 (c) we can see that for all the Fermi energies considered the Fano factor varies with $\phi $ in similar forms. When the Fermi energy $E$ and the incident angle $\theta $ are fixed, the transmission features of the conducting structure are fixed. The variation of the pumping phase lag would not change the transmission features. For all Fermi energies and incident angles, the pumping properties as functions of the driving phase difference are similar. For configurations of $E$ and $\theta $ that higher potential barriers have stronger transmission, the PC and Fano factor are positive at ${\varphi _2} - {\varphi _1} \in \left[ {\pi ,2\pi } \right]$ and negative at ${\varphi _2} - {\varphi _1} \in \left[ {0 ,\pi } \right]$. And for configurations of $E$ and $\theta $ that lower potential barriers have stronger transmission, the sign of the PC and Fano factor is reversed. At the phase lag $0$, $\pi $ , and $2 \pi $, the PC changes direction as a result of the swap of the opening order of the two gates. When the PC changes direction, interaction of electrons and holes in virtual processes is enhanced and the Fano factor demonstrates a sharp rise.

\section{Conclusions}

In summary, the PSN properties in adiabatically modulated graphene-based double-barrier structures are investigated. Within the scattering-matrix framework, general expressions for adiabatically PSN in phase-coherent
mesoscopic conductors are derived. In comparison with uncorrelated Poisson processes, numerical results of the PC, PSN, and Fano factor as functions of the incident angle, the Fermi energy of the reservoirs, and the phase difference of the two oscillating parameters are presented. It is revealed that the PSN is greatly enhanced beyond the Poisson process due to interactions of electrons and holes in Klein-type virtual tunneling processes. In particular, the PSN is
dramatically enhanced at the energy and incident angle configuration with which the dc pumped current changes flow
direction.

\section{Acknowledgements}

This project was supported by the National Natural Science
Foundation of China (No. 11004063), the Fundamental Research Funds
for the Central Universities, SCUT (No. 2009ZM0299), the Nature
Science Foundation of SCUT (No. x2lxE5090410) and the Graduate
Course Construction Project of SCUT (No. yjzk2009001 and No.
yjzk2010009).

\section{Appendix A: Derivation of the pumped shot noise}

To describe the current-current fluctuations we will use the
correlation function\cite{Ref48}
\begin{equation}
\begin{array}{l}
 S_{\alpha \beta } \left( {t,t'} \right) = \frac{1}{2}\left\langle {\Delta \hat I_\alpha  \left( t \right)\Delta \hat I_\beta  \left( {t'} \right) + \Delta \hat I_\beta  \left( {t'} \right)\Delta \hat I_\alpha  \left( t \right)} \right\rangle  \\
 \hspace{1.7cm} = \frac{1}{2}\left[ {\left\langle {\hat I_\alpha  \left( t \right)\hat I_\beta  \left( {t'} \right)} \right\rangle  + \left\langle {\hat I_\beta  \left( {t'} \right)\hat I_\alpha  \left( t \right)} \right\rangle } \right. \\
 \hspace{1.7cm} \left. { - \left\langle {\hat I_\alpha  \left( t \right)} \right\rangle \left\langle {\hat I_\beta  \left( {t'} \right)} \right\rangle  - \left\langle {\hat I_\beta  \left( {t'} \right)} \right\rangle \left\langle {\hat I_\alpha  \left( t \right)} \right\rangle } \right], \\
 \end{array}
\end{equation}
with $\Delta \hat I = \hat I - \left\langle {\hat I} \right\rangle $
and $\hat I_\alpha  \left( t \right)$ is the quantum-mechanical
current operator in the lead $\alpha $. The zero-frequency pumped
shot noise (PSN) averaged over a long time ($\Delta t \gg {{2\pi }
\mathord{\left/
 {\vphantom {{2\pi } \omega }} \right.
 \kern-\nulldelimiterspace} \omega }$) is the time integral of
 $S_{\alpha \beta } \left( {t,t'} \right)$ as follows.
\begin{equation}
S_{\alpha \beta }  = \frac{\omega }{{2\pi }}\int_{ - \infty }^{ +
\infty } {\int_0^{\frac{{2\pi }}{\omega }} {S_{\alpha \beta } \left(
{t,t'} \right)dt'dt} }
\end{equation}
The first term in the PSN is
\begin{equation}
\frac{1}{2}\frac{\omega }{{2\pi }}\int_{ - \infty }^{ + \infty }
{\int_0^{\frac{{2\pi }}{\omega }} {\left\langle {\hat I_\alpha
\left( t \right)\hat I_\beta  \left( {t'} \right)} \right\rangle
dt'dt} }
\end{equation}
with
\begin{equation}
\hat I_\alpha  \left( t \right) = \frac{e}{h}\left[ {\hat b_\alpha
^\dag  \left( t \right)\hat b_\alpha  \left( t \right) - \hat
a_\alpha ^\dag  \left( t \right)\hat a_\alpha  \left( t \right)}
\right],
\end{equation}
and
\begin{equation}
\hat I_\beta  \left( {t'} \right) = \frac{e}{h}\left[ {\hat b_\beta
^\dag  \left( {t'} \right)\hat b_\beta  \left( {t'} \right) - \hat
a_\beta ^\dag  \left( {t'} \right)\hat a_\beta  \left( {t'} \right)}
\right].
\end{equation}
Therefore, we have
\begin{equation}
\begin{array}{c}
 \hat I_\alpha  \left( t \right)\hat I_\beta  \left( {t'} \right) = \frac{{e^2 }}{{h^2 }}\left[ {\hat b_\alpha ^\dag  \left( t \right)\hat b_\alpha  \left( t \right)\hat b_\beta ^\dag  \left( {t'} \right)\hat b_\beta  \left( {t'} \right)} \right. \\
  \hspace{1.2cm} - \hat b_\alpha ^\dag  \left( t \right)\hat b_\alpha  \left( t \right)\hat a_\beta ^\dag  \left( {t'} \right)\hat a_\beta  \left( {t'} \right) \\
  \hspace{1.2cm} - \hat a_\alpha ^\dag  \left( t \right)\hat a_\alpha  \left( t \right)\hat b_\beta ^\dag  \left( {t'} \right)\hat b_\beta  \left( {t'} \right) \\
 \hspace{1.6cm} \left. { + \hat a_\alpha ^\dag  \left( t \right)\hat a_\alpha  \left( t \right)\hat a_\beta ^\dag  \left( {t'} \right)\hat a_\beta  \left( {t'} \right)} \right]. \\
 \end{array}
\end{equation}
Substituting $\hat b_\alpha  \left( t \right) =
 \sum\limits_\beta  {s_{\alpha \beta } \left( t \right)\hat a_\beta  \left( t \right)} $
 into the above equation, we have
\begin{equation}
\begin{array}{c}
 \hat I_\alpha  \left( t \right)\hat I_\beta  \left( {t'} \right) = \frac{{e^2 }}{{h^2 }}\sum\limits_{\mu \upsilon \xi \eta } {\hat a_\mu ^\dag  \left( t \right)s_{\mu \alpha }^\dag  \left( t \right)s_{\alpha \upsilon } \left( t \right)\hat a_\upsilon  \left( t \right)\hat a_\xi ^\dag  \left( {t'} \right)s_{\xi \beta }^\dag  \left( {t'} \right)s_{\beta \eta } \left( {t'} \right)\hat a_\eta  \left( {t'} \right)}  \\
  \hspace{-0.6cm} - \frac{{e^2 }}{{h^2 }}\sum\limits_{\mu \upsilon } {\hat a_\mu ^\dag  \left( t \right)s_{\mu \alpha }^\dag  \left( t \right)s_{\alpha \upsilon } \left( t \right)\hat a_\upsilon  \left( t \right)\hat a_\beta ^\dag  \left( {t'} \right)\hat a_\beta  \left( {t'} \right)}  \\
  \hspace{-0.4cm} - \frac{{e^2 }}{{h^2 }}\sum\limits_{\mu \upsilon } {\hat a_\alpha ^\dag  \left( t \right)\hat a_\alpha  \left( t \right)\hat a_\mu ^\dag  \left( {t'} \right)s_{\mu \beta }^\dag  \left( {t'} \right)s_{\beta \upsilon } \left( {t'} \right)\hat a_\upsilon  \left( {t'} \right)}  \\
  \hspace{-3.3cm} + \frac{{e^2 }}{{h^2 }}\hat a_\alpha ^\dag  \left( t \right)\hat a_\alpha  \left( t \right)\hat a_\beta ^\dag  \left( {t'} \right)\hat a_\beta  \left( {t'} \right), \\
 \end{array}
\end{equation}
and
\begin{equation}
\begin{array}{c}
 \left\langle {\hat I_\alpha  \left( t \right)} \right\rangle \left\langle {\hat I_\beta  \left( {t'} \right)} \right\rangle  = \frac{{e^2 }}{{h^2 }}\sum\limits_{\mu \upsilon \xi \eta } {\left\langle {\hat a_\mu ^\dag  \left( t \right)s_{\mu \alpha }^\dag  \left( t \right)s_{\alpha \upsilon } \left( t \right)\hat a_\upsilon  \left( t \right)} \right\rangle \left\langle {\hat a_\xi ^\dag  \left( {t'} \right)s_{\xi \beta }^\dag  \left( {t'} \right)s_{\beta \eta } \left( {t'} \right)\hat a_\eta  \left( {t'} \right)} \right\rangle }  \\
  \hspace{0.5cm} - \frac{{e^2 }}{{h^2 }}\sum\limits_{\mu \upsilon } {\left\langle {\hat a_\mu ^\dag  \left( t \right)s_{\mu \alpha }^\dag  \left( t \right)s_{\alpha \upsilon } \left( t \right)\hat a_\upsilon  \left( t \right)} \right\rangle \left\langle {\hat a_\beta ^\dag  \left( {t'} \right)\hat a_\beta  \left( {t'} \right)} \right\rangle }  \\
 \hspace{0.7cm}  - \frac{{e^2 }}{{h^2 }}\sum\limits_{\mu \upsilon } {\left\langle {\hat a_\alpha ^\dag  \left( t \right)\hat a_\alpha  \left( t \right)} \right\rangle \left\langle {\hat a_\mu ^\dag  \left( {t'} \right)s_{\mu \beta }^\dag  \left( {t'} \right)s_{\beta \upsilon } \left( {t'} \right)\hat a_\upsilon  \left( {t'} \right)} \right\rangle }  \\
  \hspace{-2.2cm} + \frac{{e^2 }}{{h^2 }}\left\langle {\hat a_\alpha ^\dag  \left( t \right)\hat a_\alpha  \left( t \right)} \right\rangle \left\langle {\hat a_\beta ^\dag  \left( {t'} \right)\hat a_\beta  \left( {t'} \right)} \right\rangle . \\
 \end{array}
\end{equation}
 Using $\hat a_\alpha  \left( t \right) = \int {dE\hat a_\alpha
\left( E \right)e^{{{ - iEt} \mathord{\left/
 {\vphantom {{ - iEt} \hbar }} \right.
 \kern-\nulldelimiterspace} \hbar }} } $ and
 $\hat a_{\alpha}^{\dag}  \left( t \right) = \int {dE\hat a_{\alpha}^{\dag}
 \left( E \right)e^{{{  iEt} \mathord{\left/
 {\vphantom {{ - iEt} \hbar }} \right.
 \kern-\nulldelimiterspace} \hbar }} } $, the first term in Eq. (19)
 reads
\begin{equation}
 \begin{array}{l}
 \frac{{e^2 }}{{h^2 }}\sum\limits_{\mu \upsilon \xi \eta } {\int {dE_1 dE_2 dE_3 dE_4 } \hat a_\mu ^\dag  \left( {E_1 } \right)e^{{{iE_1 t} \mathord{\left/
 {\vphantom {{iE_1 t} \hbar }} \right.
 \kern-\nulldelimiterspace} \hbar }} s_{\mu \alpha }^\dag  \left( t \right)s_{\alpha \upsilon } \left( t \right)\hat a_\upsilon  \left( {E_2 } \right)e^{{{ - iE_2 t} \mathord{\left/
 {\vphantom {{ - iE_2 t} \hbar }} \right.
 \kern-\nulldelimiterspace} \hbar }} }  \\
  \hspace{0.6cm} \times \hat a_\xi ^\dag  \left( {E_3 } \right)e^{{{iE_3 t'} \mathord{\left/
 {\vphantom {{iE_3 t'} \hbar }} \right.
 \kern-\nulldelimiterspace} \hbar }} s_{\xi \beta }^\dag  \left( {t'} \right)s_{\beta \eta } \left( {t'} \right)\hat a_\eta  \left( {E_4 } \right)e^{{{ - iE_4 t'} \mathord{\left/
 {\vphantom {{ - iE_4 t'} \hbar }} \right.
 \kern-\nulldelimiterspace} \hbar }} . \\
 \end{array}
\end{equation}
Wick's theorem gives the quantum statistical expectation value of
products of four operators $\hat a$. For a Fermi gas at equilibrium
this expectation value is\cite{Ref48}
\begin{equation}
\begin{array}{l}
 \left\langle {\hat a_\mu ^\dag  \left( {E_1 } \right)\hat a_\upsilon  \left( {E_2 } \right)\hat a_\xi ^\dag  \left( {E_3 } \right)\hat a_\eta  \left( {E_4 } \right)} \right\rangle  - \left\langle {\hat a_\mu ^\dag  \left( {E_1 } \right)\hat a_\upsilon  \left( {E_2 } \right)} \right\rangle \left\langle {\hat a_\xi ^\dag  \left( {E_3 } \right)\hat a_\eta  \left( {E_4 } \right)} \right\rangle  \\
  = \delta _{\mu \eta } \delta _{\upsilon \xi } \delta \left( {E_1  - E_4 } \right)\delta \left( {E_2  - E_3 } \right)f_\mu  \left( {E_1 } \right)\left[ {1 - f_\upsilon  \left( {E_2 } \right)} \right]. \\
 \end{array}
\end{equation}
$f_{\alpha } (E)$ is the Fermi distribution function of the $\alpha
$ reservoir connected to the adiabatically modulated conductor.
Substituting Eq. (22) into the first term of $\left\langle {\hat
I_\alpha \left( t \right)\hat I_\beta  \left( {t'} \right)}
\right\rangle  - \left\langle {\hat I_\alpha  \left( t \right)}
\right\rangle \left\langle {\hat I_\beta  \left( {t'} \right)}
\right\rangle $, we have
\begin{equation}
\begin{array}{l}
 \frac{{e^2 }}{{h^2 }}\sum\limits_{\mu \upsilon \xi \eta } {\int {dE_1 dE_2 dE_3 dE_4 } \delta _{\mu \eta } \delta _{\nu \xi } \delta \left( {E_1  - E_4 } \right)\delta \left( {E_2  - E_3 } \right)f_\mu  \left( {E_1 } \right)\left[ {1 - f_\nu  \left( {E_2 } \right)} \right]}  \\
  \hspace{0.6cm} \times e^{{{iE_1 t} \mathord{\left/
 {\vphantom {{iE_1 t} \hbar }} \right.
 \kern-\nulldelimiterspace} \hbar }} s_{\mu \alpha }^\dag  \left( t \right)s_{\alpha \upsilon } \left( t \right)e^{{{ - iE_2 t} \mathord{\left/
 {\vphantom {{ - iE_2 t} \hbar }} \right.
 \kern-\nulldelimiterspace} \hbar }} e^{{{iE_3 t'} \mathord{\left/
 {\vphantom {{iE_3 t'} \hbar }} \right.
 \kern-\nulldelimiterspace} \hbar }} s_{\xi \beta }^\dag  \left( {t'} \right)s_{\beta \eta } \left( {t'} \right)e^{{{ - iE_4 t'} \mathord{\left/
 {\vphantom {{ - iE_4 t'} \hbar }} \right.
 \kern-\nulldelimiterspace} \hbar }} . \\
 \end{array}
\end{equation}
Integrating out $\eta $, $\xi $, $E_4$, and $E_3$, we obtain
\begin{equation}
\begin{array}{l}
 \frac{{e^2 }}{{h^2 }}\sum\limits_{\mu \upsilon } {\int {dE_1 dE_2 } f_\mu  \left( {E_1 } \right)\left[ {1 - f_\nu  \left( {E_2 } \right)} \right]e^{{{iE_1 t} \mathord{\left/
 {\vphantom {{iE_1 t} \hbar }} \right.
 \kern-\nulldelimiterspace} \hbar }} s_{\mu \alpha }^\dag  \left( t \right)}  \\
  \hspace{0.6cm} \times s_{\alpha \upsilon } \left( t \right)e^{{{ - iE_2 t} \mathord{\left/
 {\vphantom {{ - iE_2 t} \hbar }} \right.
 \kern-\nulldelimiterspace} \hbar }} e^{{{iE_2 t'} \mathord{\left/
 {\vphantom {{iE_2 t'} \hbar }} \right.
 \kern-\nulldelimiterspace} \hbar }} s_{\nu \beta }^\dag  \left( {t'} \right)s_{\beta \mu } \left( {t'} \right)e^{{{ - iE_1 t'} \mathord{\left/
 {\vphantom {{ - iE_1 t'} \hbar }} \right.
 \kern-\nulldelimiterspace} \hbar }} . \\
 \end{array}
\end{equation}
Following similar procedures to all the other terms in Eq. (13), we
can obtain
\begin{equation}
\begin{array}{l}
 S_{\alpha \beta } \left( {t,t'} \right) = \frac{{e^2 }}{{2h^2 }}\sum\limits_{\mu \upsilon } {\left[ {\int {dE_1 dE_2 } f_\mu  \left( {E_1 } \right)\left[ {1 - f_\nu  \left( {E_2 } \right)} \right]e^{{{iE_1 t} \mathord{\left/
 {\vphantom {{iE_1 t} \hbar }} \right.
 \kern-\nulldelimiterspace} \hbar }} s_{\mu \alpha }^\dag  \left( t \right)} \right.}  \\
 \left. { \times s_{\alpha \upsilon } \left( t \right)e^{{{ - iE_2 t} \mathord{\left/
 {\vphantom {{ - iE_2 t} \hbar }} \right.
 \kern-\nulldelimiterspace} \hbar }} e^{{{iE_2 t'} \mathord{\left/
 {\vphantom {{iE_2 t'} \hbar }} \right.
 \kern-\nulldelimiterspace} \hbar }} s_{\nu \beta }^\dag  \left( {t'} \right)s_{\beta \mu } \left( {t'} \right)e^{{{ - iE_1 t'} \mathord{\left/
 {\vphantom {{ - iE_1 t'} \hbar }} \right.
 \kern-\nulldelimiterspace} \hbar }} } \right] \\
  - \frac{{e^2 }}{{2h^2 }}\int {dE_1 dE_2 } f_\beta  \left( {E_1 } \right)\left[ {1 - f_\beta  \left( {E_2 } \right)} \right]e^{{{iE_1 t} \mathord{\left/
 {\vphantom {{iE_1 t} \hbar }} \right.
 \kern-\nulldelimiterspace} \hbar }} s_{\beta \alpha }^\dag  \left( t \right)s_{\alpha \beta } \left( t \right)e^{{{ - iE_2 t} \mathord{\left/
 {\vphantom {{ - iE_2 t} \hbar }} \right.
 \kern-\nulldelimiterspace} \hbar }} e^{{{iE_2 t'} \mathord{\left/
 {\vphantom {{iE_2 t'} \hbar }} \right.
 \kern-\nulldelimiterspace} \hbar }} e^{{{ - iE_1 t'} \mathord{\left/
 {\vphantom {{ - iE_1 t'} \hbar }} \right.
 \kern-\nulldelimiterspace} \hbar }}  \\
  - \frac{{e^2 }}{{2h^2 }}\int {dE_1 dE_2 } f_\alpha  \left( {E_1 } \right)\left[ {1 - f_\alpha  \left( {E_2 } \right)} \right]e^{{{iE_1 t} \mathord{\left/
 {\vphantom {{iE_1 t} \hbar }} \right.
 \kern-\nulldelimiterspace} \hbar }} e^{{{ - iE_2 t} \mathord{\left/
 {\vphantom {{ - iE_2 t} \hbar }} \right.
 \kern-\nulldelimiterspace} \hbar }} e^{{{iE_2 t'} \mathord{\left/
 {\vphantom {{iE_2 t'} \hbar }} \right.
 \kern-\nulldelimiterspace} \hbar }} s_{\alpha \beta }^\dag  \left( {t'} \right)s_{\beta \alpha } \left( {t'} \right)e^{{{ - iE_1 t'} \mathord{\left/
 {\vphantom {{ - iE_1 t'} \hbar }} \right.
 \kern-\nulldelimiterspace} \hbar }}  \\
  + \frac{{e^2 }}{{2h^2 }}\int {dE_1 dE_2 } f_\alpha  \left( {E_1 } \right)\left[ {1 - f_\alpha  \left( {E_2 } \right)} \right]e^{{{iE_1 t} \mathord{\left/
 {\vphantom {{iE_1 t} \hbar }} \right.
 \kern-\nulldelimiterspace} \hbar }} e^{{{ - iE_2 t} \mathord{\left/
 {\vphantom {{ - iE_2 t} \hbar }} \right.
 \kern-\nulldelimiterspace} \hbar }} e^{{{iE_2 t'} \mathord{\left/
 {\vphantom {{iE_2 t'} \hbar }} \right.
 \kern-\nulldelimiterspace} \hbar }} e^{{{ - iE_1 t'} \mathord{\left/
 {\vphantom {{ - iE_1 t'} \hbar }} \right.
 \kern-\nulldelimiterspace} \hbar }}  \\
  + \frac{{e^2 }}{{2h^2 }}\sum\limits_{\mu \upsilon } {\left[ {\int {dE_1 dE_2 } f_\mu  \left( {E_1 } \right)\left[ {1 - f_\nu  \left( {E_2 } \right)} \right]e^{{{iE_1 t'} \mathord{\left/
 {\vphantom {{iE_1 t'} \hbar }} \right.
 \kern-\nulldelimiterspace} \hbar }} s_{\mu \beta }^\dag  \left( {t'} \right)} \right.}  \\
 \left. { \times s_{\beta \upsilon } \left( {t'} \right)e^{{{ - iE_2 t'} \mathord{\left/
 {\vphantom {{ - iE_2 t'} \hbar }} \right.
 \kern-\nulldelimiterspace} \hbar }} e^{{{iE_2 t} \mathord{\left/
 {\vphantom {{iE_2 t} \hbar }} \right.
 \kern-\nulldelimiterspace} \hbar }} s_{\nu \alpha }^\dag  \left( t \right)s_{\alpha \mu } \left( t \right)e^{{{ - iE_1 t} \mathord{\left/
 {\vphantom {{ - iE_1 t} \hbar }} \right.
 \kern-\nulldelimiterspace} \hbar }} } \right] \\
  - \frac{{e^2 }}{{2h^2 }}\int {dE_1 dE_2 } f_\alpha  \left( {E_1 } \right)\left[ {1 - f_\alpha  \left( {E_2 } \right)} \right]e^{{{iE_1 t'} \mathord{\left/
 {\vphantom {{iE_1 t'} \hbar }} \right.
 \kern-\nulldelimiterspace} \hbar }} s_{\alpha \beta }^\dag  \left( {t'} \right)s_{\beta \alpha } \left( {t'} \right)e^{{{ - iE_2 t'} \mathord{\left/
 {\vphantom {{ - iE_2 t'} \hbar }} \right.
 \kern-\nulldelimiterspace} \hbar }} e^{{{iE_2 t} \mathord{\left/
 {\vphantom {{iE_2 t} \hbar }} \right.
 \kern-\nulldelimiterspace} \hbar }} e^{{{ - iE_1 t} \mathord{\left/
 {\vphantom {{ - iE_1 t} \hbar }} \right.
 \kern-\nulldelimiterspace} \hbar }}  \\
  - \frac{{e^2 }}{{2h^2 }}\int {dE_1 dE_2 } f_\beta  \left( {E_1 } \right)\left[ {1 - f_\beta  \left( {E_2 } \right)} \right]e^{{{iE_1 t'} \mathord{\left/
 {\vphantom {{iE_1 t'} \hbar }} \right.
 \kern-\nulldelimiterspace} \hbar }} e^{{{ - iE_2 t'} \mathord{\left/
 {\vphantom {{ - iE_2 t'} \hbar }} \right.
 \kern-\nulldelimiterspace} \hbar }} e^{{{iE_2 t} \mathord{\left/
 {\vphantom {{iE_2 t} \hbar }} \right.
 \kern-\nulldelimiterspace} \hbar }} s_{\beta \alpha }^\dag  \left( t \right)s_{\alpha \beta } \left( t \right)e^{{{ - iE_1 t} \mathord{\left/
 {\vphantom {{ - iE_1 t} \hbar }} \right.
 \kern-\nulldelimiterspace} \hbar }}  \\
  + \frac{{e^2 }}{{2h^2 }}\int {dE_1 dE_2 } f_\beta  \left( {E_1 } \right)\left[ {1 - f_\beta  \left( {E_2 } \right)} \right]e^{{{iE_1 t'} \mathord{\left/
 {\vphantom {{iE_1 t'} \hbar }} \right.
 \kern-\nulldelimiterspace} \hbar }} e^{{{ - iE_2 t'} \mathord{\left/
 {\vphantom {{ - iE_2 t'} \hbar }} \right.
 \kern-\nulldelimiterspace} \hbar }} e^{{{iE_2 t} \mathord{\left/
 {\vphantom {{iE_2 t} \hbar }} \right.
 \kern-\nulldelimiterspace} \hbar }} e^{{{ - iE_1 t} \mathord{\left/
 {\vphantom {{ - iE_1 t} \hbar }} \right.
 \kern-\nulldelimiterspace} \hbar }} . \\
 \end{array}
\end{equation}
The first term of the above equation has a product of four
scattering matrix elements. We list the four scattering matrix
expanded into the form of Eq. (6) as
\begin{equation}
\begin{array}{*{20}c}
   1 & 2 & 3  \\
   {\left( {s_{\mu \alpha }^{\dag 0} } \right.} & { + s_{\mu \alpha }^{\dag  - \omega } e^{ - i\omega t} } & {\left. { + s_{\mu \alpha }^{\dag  + \omega } e^{i\omega t} } \right)}  \\
   {\left( {s_{\alpha \nu }^0 } \right.} & { + s_{\alpha \nu }^{ - \omega } e^{i\omega t} } & {\left. { + s_{\alpha \nu }^{ + \omega } e^{ - i\omega t} } \right)}  \\
   {\left( {s_{\nu \beta }^{\dag 0} } \right.} & { + s_{\nu \beta }^{\dag  - \omega } e^{ - i\omega t'} } & {\left. { + s_{\nu \beta }^{\dag  + \omega } e^{i\omega t'} } \right)}  \\
   {\left( {s_{\beta \mu }^0 } \right.} & { + s_{\beta \mu }^{ - \omega } e^{i\omega t'} } & {\left. { + s_{\beta \mu }^{ + \omega } e^{ - i\omega t'} } \right)}
   .\\
\end{array}
\end{equation}
We calculate the column 1111 term of Eq. (26) in the time-averaged
zero-frequency PSN as
\begin{equation}
\begin{array}{l}
 \frac{{e^2 }}{{h^2 }}\frac{\omega }{{4\pi }}\sum\limits_{\mu \nu } {\int {dE_1 dE_2 } \int_{ - \infty }^{ + \infty } {dt} \int_0^{\frac{{2\pi }}{\omega }} {dt'\left[ {f_\mu  \left( {E_1 } \right)\left[ {1 - f_\upsilon  \left( {E_2 } \right)} \right]e^{{{iE_1 t} \mathord{\left/
 {\vphantom {{iE_1 t} \hbar }} \right.
 \kern-\nulldelimiterspace} \hbar }} } \right.} }  \\
 \hspace{1cm} \left. { \times s_{\mu \alpha }^{\dag 0} s_{\alpha \upsilon }^0 s_{\upsilon \beta }^{\dag 0} s_{\beta \mu }^0 e^{{{ - iE_2 t} \mathord{\left/
 {\vphantom {{ - iE_2 t} \hbar }} \right.
 \kern-\nulldelimiterspace} \hbar }} e^{{{iE_2 t'} \mathord{\left/
 {\vphantom {{iE_2 t'} \hbar }} \right.
 \kern-\nulldelimiterspace} \hbar }} e^{{{ - iE_1 t'} \mathord{\left/
 {\vphantom {{ - iE_1 t'} \hbar }} \right.
 \kern-\nulldelimiterspace} \hbar }} } \right]. \\
 \end{array}
\end{equation}
From the relation $\frac{1}{{2\pi }} \int_{ - \infty }^{ + \infty }
{dte^{{{i\left( {E_1 - E_2 } \right)t} \mathord{\left/
 {\vphantom {{i\left( {E_1  - E_2 } \right)t} \hbar }} \right.
 \kern-\nulldelimiterspace} \hbar }} }  =\hbar \delta \left( {E_1  - E_2 }
 \right)$, it can be seen that the two-fold integral over the energy is
 reduced to one. For the configuration of a quantum pump, no bias is
 applied. Therefore for any value of the energy, the Fermi
 distribution function $f_\alpha  \left( E \right)$ is
 simultaneously $1$ or $0$ at zero temperature for all leads. Hence,
$ {f_\mu  \left( {E } \right)\left[ {1 - f_\nu  \left( {E } \right)}
\right]=0} $ for any $\mu$ and $\nu$s. We can achieve that Eq. (27)
is equal to zero. For the same reason, all the 11** term taken into
the PSN are equal to zero since the $t'$ exponential $e^{\pm i
\omega t'}$ would not affect the integral of the time $t$. Then we
go to the 1211 term taken into the time-averaged zero-frequency PSN:
\begin{equation}
\begin{array}{l}
 \frac{{e^2 }}{{h^2 }}\frac{\omega }{{4\pi }}\sum\limits_{\mu \nu } {\int {dE_1 dE_2 } \int_{ - \infty }^{ + \infty } {dt} \int_0^{\frac{{2\pi }}{\omega }} {dt'\left[ {f_\mu  \left( {E_1 } \right)\left[ {1 - f_\upsilon  \left( {E_2 } \right)} \right]e^{{{iE_1 t} \mathord{\left/
 {\vphantom {{iE_1 t} \hbar }} \right.
 \kern-\nulldelimiterspace} \hbar }} } \right.} }  \\
 \hspace{1cm} \left. { \times s_{\mu \alpha }^{\dag 0} s_{\alpha \upsilon }^{ - \omega } e^{i\omega t} s_{\upsilon \beta }^{\dag 0} s_{\beta \mu }^0 e^{{{ - iE_2 t} \mathord{\left/
 {\vphantom {{ - iE_2 t} \hbar }} \right.
 \kern-\nulldelimiterspace} \hbar }} e^{{{iE_2 t'} \mathord{\left/
 {\vphantom {{iE_2 t'} \hbar }} \right.
 \kern-\nulldelimiterspace} \hbar }} e^{{{ - iE_1 t'} \mathord{\left/
 {\vphantom {{ - iE_1 t'} \hbar }} \right.
 \kern-\nulldelimiterspace} \hbar }} } \right]. \\
 \end{array}
\end{equation}
With the definition of the $\delta $ function
\begin{equation}
\frac{1}{{2\pi }} \int_{ - \infty }^{ + \infty } {dte^{{{i\left(
{E_1  + \hbar \omega - E_2 } \right)t} \mathord{\left/
 {\vphantom {{i\left( {E_1  + \hbar \omega  - E_2 } \right)t} \hbar }} \right.
 \kern-\nulldelimiterspace} \hbar }} }  = \hbar \delta \left( {E_1  + \hbar \omega  - E_2 }
 \right),
\end{equation}
we get
\begin{equation}
\frac{{e^2 }}{h}\frac{\omega }{{4 \pi  }}\sum\limits_{\mu \nu }
{\int {dE_1 } \int_0^{\frac{{2\pi }}{\omega }} {dt'f_\mu  \left(
{E_1 } \right)\left[ {1 - f_\upsilon  \left( {E_1  + \hbar \omega }
\right)} \right]s_{\mu \alpha }^{\dag 0} s_{\alpha \upsilon }^{ -
\omega } s_{\upsilon \beta }^{\dag 0} s_{\beta \mu }^0 e^{i\omega
t'} } } .
\end{equation}
$e^{i\omega t'}$ is a periodic function of $t'$ with the period
$2\pi / \omega$. Its integral over one period is zero. Therefore the
above whole term is zero. Similarly, the 1212 term is zero with an
additional exponential $e^{i \omega t'}$ the only difference from
the 1211 term, whose one-period-integral is again zero. Following
analogous procedures, we can derive the 1213 term as
\begin{equation}
\begin{array}{l}
 \frac{{e^2 }}{h}\frac{\omega }{{4\pi }}\sum\limits_{\mu \nu } {\int {dE_1 } \int_0^{\frac{{2\pi }}{\omega }} {dt'f_\mu  \left( {E_1 } \right)\left[ {1 - f_\upsilon  \left( {E_1  + \hbar \omega } \right)} \right]s_{\mu \alpha }^{\dag 0} s_{\alpha \upsilon }^{ - \omega } s_{\upsilon \beta }^{\dag 0} s_{\beta \mu }^{ + \omega } } }  \\
  \hspace{1cm} = \frac{{e^2 }}{{2h}}\sum\limits_{\mu \nu } {\int {dE_1 } f_\mu  \left( {E_1 } \right)\left[ {1 - f_\upsilon  \left( {E_1  + \hbar \omega } \right)} \right]s_{\mu \alpha }^{\dag 0} s_{\alpha \upsilon }^{ - \omega } s_{\upsilon \beta }^{\dag 0} s_{\beta \mu }^{ + \omega } } . \\
 \end{array}
\end{equation}
The quantum pumping configuration sets equal chemical potentials in
all reservoirs, i.e., for any $\alpha $, we have
\begin{equation}
f_\alpha  \left( E \right) = \left\{ \begin{array}{l}
 \begin{array}{*{20}c}
   {1,} & {E \le \mu ,}  \\
\end{array} \\
 \begin{array}{*{20}c}
   {0,} & {E > \mu .}  \\
\end{array} \\
 \end{array} \right.
\end{equation}
Hence, only the integral range $\int_{\mu  - \hbar \omega }^\mu
{dE_1 } $ contributes in Eq. (31), which is
\begin{equation}
\frac{{e^2 \omega }}{{4\pi }}\sum\limits_{\mu \nu } {s_{\mu \alpha
}^{\dag 0} s_{\alpha \upsilon }^{ - \omega } s_{\upsilon \beta
}^{\dag 0} s_{\beta \mu }^{ + \omega } }.
\end{equation}
Analogously, the 1221 term is equal to
\begin{equation}
\frac{{e^2 \omega }}{{4\pi }}\sum\limits_{\mu \nu } {s_{\mu \alpha
}^{\dag 0} s_{\alpha \upsilon }^{ - \omega } s_{\upsilon \beta
}^{\dag  - \omega } s_{\beta \mu }^0 }.
\end{equation}
Following similar algebra, we could see that the 1222, 1223, $
\cdots $ , 3111, 3112 terms are all zero. And the 3113 term is equal
to
\begin{equation}
\frac{{e^2 \omega }}{{4\pi }}\sum\limits_{\mu \nu } {s_{\mu \alpha
}^{\dag  + \omega } s_{\alpha \upsilon }^0 s_{\upsilon \beta }^{\dag
0} s_{\beta \mu }^{ + \omega } } .
\end{equation}
The 3121 term is equal to
\begin{equation}
\frac{{e^2 \omega }}{{4\pi }}\sum\limits_{\mu \nu } {s_{\mu \alpha
}^{\dag  + \omega } s_{\alpha \upsilon }^0 s_{\upsilon \beta }^{\dag
- \omega } s_{\beta \mu }^0 } .
\end{equation}
The 3122, 3123, $ \cdots $ , 3221, 3222 terms are all zero. The 3223
term is equal to
\begin{equation}
\begin{array}{l}
 \frac{{e^2 }}{h}\frac{\omega }{{4\pi }}\sum\limits_{\mu \nu } {\int {dE_1 } \int_0^{\frac{{2\pi }}{\omega }} {dt'f_\mu  \left( {E_1 } \right)\left[ {1 - f_\upsilon  \left( {E_1  + 2\hbar \omega } \right)} \right]} }  \\
  \hspace{2cm} \times s_{\mu \alpha }^{\dag  + \omega } s_{\alpha \upsilon }^{ - \omega } s_{\upsilon \beta }^{\dag  - \omega } s_{\beta \mu }^{ + \omega }  \\
 \hspace{1cm} = \frac{{e^2 \omega }}{{2\pi }}\sum\limits_{\mu \nu } {s_{\mu \alpha }^{\dag 0} s_{\alpha \upsilon }^{ - \omega } s_{\upsilon \beta }^{\dag 0} s_{\beta \mu }^{ + \omega } } . \\
 \end{array}
\end{equation}
The rest terms from 3231 to 3333 are all zero. Following similar
algebra, we could obtain that the two-scattering-matrix and
no-scattering-matrix terms are all equal to zero. And the
contribution of $\left\langle {\hat I_\beta  \left( {t'} \right)\hat
I_\alpha  \left( t \right)} \right\rangle  - \left\langle {\hat
I_\beta  \left( {t'} \right)} \right\rangle \left\langle {\hat
I_\alpha  \left( t \right)} \right\rangle $ follows from that of
$\left\langle {\hat I_\alpha  \left( t \right)\hat I_\beta  \left(
{t'} \right)} \right\rangle  - \left\langle {\hat I_\alpha  \left( t
\right)} \right\rangle \left\langle {\hat I_\beta  \left( {t'}
\right)} \right\rangle $. Totally five plus five terms contribute to
the time-averaged zero-frequency PSN. Collecting the above results
and using the expansion of the scattering matrix [Eqs. (6) and (7)],
we reach the general expression of the time-averaged zero-frequency
PSN.
\begin{equation}
\begin{array}{c}
 S_{\alpha \beta }  = \frac{{e^2 \omega }}{{2\pi }}\sum\limits_{\mu \nu j_1 j_2 } {X_{\omega ,j_2 } X_{\omega ,j_1 } s_{\upsilon \beta }^{\dag 0} \frac{{\partial s_{\alpha \nu } }}{{\partial X_{j_1 } }}\frac{{\partial s_{\beta \mu } }}{{\partial X_{j_2 } }}s_{\mu \alpha }^{\dag 0} \cos \left( {\varphi _{j_1 }  - \varphi _{j_2 } } \right)}  \\
  \hspace{0.8cm} + \frac{{e^2 \omega }}{{2\pi }}\sum\limits_{\mu \nu j_1 j_2 } {X_{\omega ,j_2 } X_{\omega ,j_1 } s_{\upsilon \beta }^{\dag 0} s_{\alpha \nu }^0 \frac{{\partial s_{\beta \mu } }}{{\partial X_{j_2 } }}\frac{{\partial s_{\mu \alpha }^\dag  }}{{\partial X_{j_1 } }}\cos \left( {\varphi _{j_1 }  - \varphi _{j_2 } } \right)}  \\
  \hspace{0.8cm} + \frac{{e^2 \omega }}{{2\pi }}\sum\limits_{\mu \nu j_1 j_2 } {X_{\omega ,j_2 } X_{\omega ,j_1 } \frac{{\partial s_{\upsilon \beta }^\dag  }}{{\partial X_{j_2 } }}\frac{{\partial s_{\alpha \upsilon } }}{{\partial X_{j_1 } }}s_{\beta \mu }^0 s_{\mu \alpha }^{\dag 0} \cos \left( {\varphi _{j_1 }  - \varphi _{j_2 } } \right)}  \\
  \hspace{0.8cm} + \frac{{e^2 \omega }}{{2\pi }}\sum\limits_{\mu \nu j_1 j_2 } {X_{\omega ,j_2 } X_{\omega ,j_1 } \frac{{\partial s_{\upsilon \beta }^\dag  }}{{\partial X_{j_2 } }}s_{\alpha \upsilon }^0 s_{\beta \mu }^0 \frac{{\partial s_{\mu \alpha }^\dag  }}{{\partial X_{j_1 } }}\cos \left( {\varphi _{j_1 }  - \varphi _{j_2 } } \right)}  \\
  \hspace{-0.4cm} + \frac{{e^2 \omega }}{{2\pi }}\sum\limits_{\mu \upsilon j_1 j_2 j_3 j_4 } {\left[ {X_{\omega ,j_1 } X_{\omega ,j_4 } X_{\omega ,j_2 } X_{\omega ,j_3 } \frac{{\partial s_{\beta \mu } }}{{\partial X_{j_4 } }}\frac{{\partial s_{\mu \alpha }^\dag  }}{{\partial X_{j_1 } }}} \right.}  \\
 \hspace{0.8cm} \left. { \times \frac{{\partial s_{\alpha \upsilon } }}{{\partial X_{j_2 } }}\frac{{\partial s_{\upsilon \beta }^\dag  }}{{\partial X_{j_3 } }}\cos \left( {\varphi _{j_4 }  - \varphi _{j_1 }  + \varphi _{j_3 }  - \varphi _{j_2 } } \right)} \right]. \\
 \end{array}
\end{equation}

\section{Appendix B: Discussion of the Poissonian pumped shot noise}

The Schottky's result\cite{Ref48, Ref49} for the shot noise corresponds to the uncorrelated arrival of particles with a distribution function of
time intervals between arrival times which is Poissonian, $P\left( {\Delta t} \right) = {\tau ^{ - 1}}\exp \left( { - {{\Delta t} \mathord{\left/
 {\vphantom {{\Delta t} \tau }} \right.
 \kern-\nulldelimiterspace} \tau }} \right)$ with $\tau $ being the
mean time interval between carriers. [$P\left( {\Delta t} \right)$ is normalized  with $\int_0^{ + \infty } {P\left( {\Delta t} \right)d\left( {\Delta t} \right)}  = 1$ and $\int_0^{ + \infty } {\left( {\Delta t} \right)P\left( {\Delta t} \right)d\left( {\Delta t} \right)}  = \tau $]. With the Poissonian time interval distribution function, we could consider the Poissonian current and shot noise. It is convenient to look at a single-electron tunneling process with $P\left( {\Delta t} \right)$ normalized to $1$ and the complete relevant time range is in the order of $\tau $.

We take an infinitesimal time segment $\left[ {t,t + dt} \right]$ from the continuous time flow in $[0, + \infty )$. The time dependent current generated by the reservoir could be expressed as
\begin{equation}
I\left( t \right) = \frac{{\int_t^{t + dt} {eP\left( {t'} \right)dt'} }}{{dt}} = \frac{e}{\tau }{e^{ - {t \mathord{\left/
 {\vphantom {t \tau }} \right.
 \kern-\nulldelimiterspace} \tau }}}.
\end{equation}
The mean current follows as
\begin{equation}
\overline {I\left( t \right)}  = \mathop {\lim }\limits_{T \to \infty } \frac{1}{T}\int_0^T {I\left( t \right)dt}  = \frac{1}{\tau }\int_0^{ + \infty } {I\left( t \right)dt}  = \frac{e}{\tau }.
\end{equation}
Here the single-electron-tunneling picture is used. The mathematical object which allows us to characterize the duration of the current pulse is called the autocorrelation function and is defined by
\begin{equation}
{R_I}\left( {t'} \right) = \mathop {\lim }\limits_{T \to \infty } \frac{1}{T}\int_{ - {T \mathord{\left/
 {\vphantom {T 2}} \right.
 \kern-\nulldelimiterspace} 2}}^{{T \mathord{\left/
 {\vphantom {T 2}} \right.
 \kern-\nulldelimiterspace} 2}} {I\left( t \right)I\left( {t + t'} \right)dt} .
\end{equation}
From the time-dependent current, we can obtain the autocorrelation function as
\begin{equation}
{R_I}\left( {t'} \right) = {\left. {\overline {I\left( t \right)I\left( {t + t'} \right)} } \right|_t} = {\left. {\overline {\frac{{{e^2}}}{{{\tau ^2}}}{e^{ - \frac{{2t}}{\tau }}}} } \right|_t}{e^{ - \frac{{t'}}{\tau }}}.
\end{equation}
The footnote $t$ means the mean value is evaluated relative to the variable $t$.
Using the following relation coming from the result of Eq. (40)
\begin{equation}
{\left. {\overline {\frac{e}{\tau }{e^{ - \frac{{2t}}{\tau }}}} } \right|_t} = {\left. {\overline {\frac{1}{2}\frac{e}{{\frac{\tau }{2}}}{e^{ - \frac{t}{{\frac{\tau }{2}}}}}} } \right|_t} = \frac{1}{2}\frac{e}{{\frac{\tau }{2}}} = \frac{e}{\tau },
\end{equation}
we have
\begin{equation}
{R_I}\left( {t'} \right) = \frac{{{e^2}}}{{{\tau ^2}}}{e^{ - \frac{{t'}}{\tau }}}.
\end{equation}
The Wiener-Khinchin theorem states that the noise spectrum is the Fourier transform of the autocorrelation function:
\begin{equation}
{S_I}\left( f \right) = 2\int_0^\infty  {{R_I}\left( {t'} \right){e^{ - i2\pi ft'}}dt'} .
\end{equation}
Therefore, the zero-frequency shot noise
\begin{equation}
{S_I}\left( 0 \right) = 2\int_0^{  \infty } {\frac{{{e^2}}}{{{\tau ^2}}}{e^{ - \frac{{t'}}{\tau }}}dt'}  = 2\frac{{{e^2}}}{\tau } = 2e\overline I ,
\end{equation}
which is just the Poisson shot noise.

Following that, we consider the pumping configuration to achieve the poissonian quantum pumped shot noise. To achieve a pure poisson process, we should exclude all conducting structure and let the conductance totally governed by two Poisson-distributed random emitters at the left and right leads since any scattering structure would induce interactions and break the Poissonian picture. The pumping mechanism is thus reduced to a semi-classical one with two modulating gates and a single-particle level between the two gates. The two gates are modulated with a phase lag $\phi  = {\pi  \mathord{\left/
 {\vphantom {\pi  2}} \right.
 \kern-\nulldelimiterspace} 2}$. We assume the gates to be two oscillating semi-classical potential barrier with the time dependence of their heights as follows.
 \begin{equation}
 \left\{ \begin{array}{l}
{U_1} = \sin \left( {t + \frac{\pi }{2}} \right),\\
{U_2} = \sin \left( t \right).
\end{array} \right.
 \end{equation}
In typical quantum pumps, the oscillation period $T = {{2\pi } \mathord{\left/
 {\vphantom {{2\pi } \omega }} \right.
 \kern-\nulldelimiterspace} \omega }$ is much larger than the mean time interval between carriers $\tau $. Here the pumping frequency $\omega $ is set to be $1$ without blurring any physics. We divide one pumping period into four quarters. When $t \in \left[ {0,{\pi  \mathord{\left/
 {\vphantom {\pi  2}} \right.
 \kern-\nulldelimiterspace} 2}} \right]$, $\sin \left( t \right)$ changes from 0 to 1 and
$\sin \left( {t + {\pi  \mathord{\left/
 {\vphantom {\pi  2}} \right.
 \kern-\nulldelimiterspace} 2}} \right)$ changes from 1 to 0. Considering the integral
effect, the two gates are equally high and the system could be approximated by two identical emitter shooting electrons at each other with a possible emission phase lag. The time-dependent current could be formulated as
\begin{equation}
I_{p}\left( t \right) = \frac{e}{\tau }{e^{\frac{{t - {t_{0L}}}}{\tau }}} - \frac{e}{\tau }{e^{\frac{{t - {t_{0R}}}}{\tau }}}.
\end{equation}
For two uncorrelated emitter, $t_{0L}$ and $t_{0R}$ are possibly different.
When $t \in \left[ {{\pi  \mathord{\left/
 {\vphantom {\pi  2}} \right.
 \kern-\nulldelimiterspace} 2},\pi } \right]$, $\sin \left( t \right)$
changes from 1 to 0 and $\sin \left( {t + {\pi  \mathord{\left/
 {\vphantom {\pi  2}} \right.
 \kern-\nulldelimiterspace} 2}} \right)$ changes from 0 to -1. In this quarter, the gate $U_{1}$ is open and the gate $U_{2}$ is closed. The electron has some probability to be emitted from the left reservoir to the middle single-electron level and fill it. There is a current flow from the left reservoir to the middle level. The time-dependent current flow from the left emitter to the middle level could be formulated as
 \begin{equation}
 {I_p}\left( t \right) = \frac{e}{\tau }{e^{\frac{{t - t{'_{0L}}}}{\tau }}}.
\end{equation}
When $t \in \left[ {\pi ,{{3\pi } \mathord{\left/
 {\vphantom {{3\pi } 2}} \right.
 \kern-\nulldelimiterspace} 2}} \right]$, $\sin \left( t \right)$
changes from 0 to -1 and $\sin \left( {t + {\pi  \mathord{\left/
 {\vphantom {\pi  2}} \right.
 \kern-\nulldelimiterspace} 2}} \right)$ changes from -1 to
0. The integral effects of the two gates balance out. The electron could not tunnel out of the middle level. When $t \in \left[ {{{3\pi } \mathord{\left/
 {\vphantom {{3\pi } 2}} \right.
 \kern-\nulldelimiterspace} 2},2\pi } \right]$, $\sin \left( t \right)$ changes from -1 to 0 and $\sin \left( {t + {\pi  \mathord{\left/
 {\vphantom {\pi  2}} \right.
 \kern-\nulldelimiterspace} 2}} \right)$
changes from 0 to 1. $U_1$ maintains higher than $U_2$.
The left gate is closed and the right gate is open, which drives the particle in the middle level to the right reservoir. As the right reservoir is a Poisson source and simultaneously a Poisson drain, the tunneling from the middle level would also be time-dependent as
\begin{equation}
{I_p}\left( t \right) = \frac{e}{\tau }{e^{\frac{{t - t{'_{0R}}}}{\tau }}}.
\end{equation}
For adiabatic quantum pumps, ${T \mathord{\left/
 {\vphantom {T 4}} \right.
 \kern-\nulldelimiterspace} 4} \gg \tau $. Therefore, the time average in one period could be approximated as the time average in the infinite time interval $[0, + \infty )$. Following similar derivation as the ordinary conductor, we could obtain
 \begin{equation}
 \overline {{I_p}\left( t \right)}  = \frac{e}{\tau }.
\end{equation}
And the the zero-frequency shot noise
\begin{equation}
{S_p}\left( 0 \right) = 2\frac{{{e^2}}}{\tau } = 2e\overline {{I_p}} ,
\end{equation}
which is the Poisson pumped shot noise.

\clearpage

\begin{figure}[h]
\includegraphics[height=14cm, width=10cm]{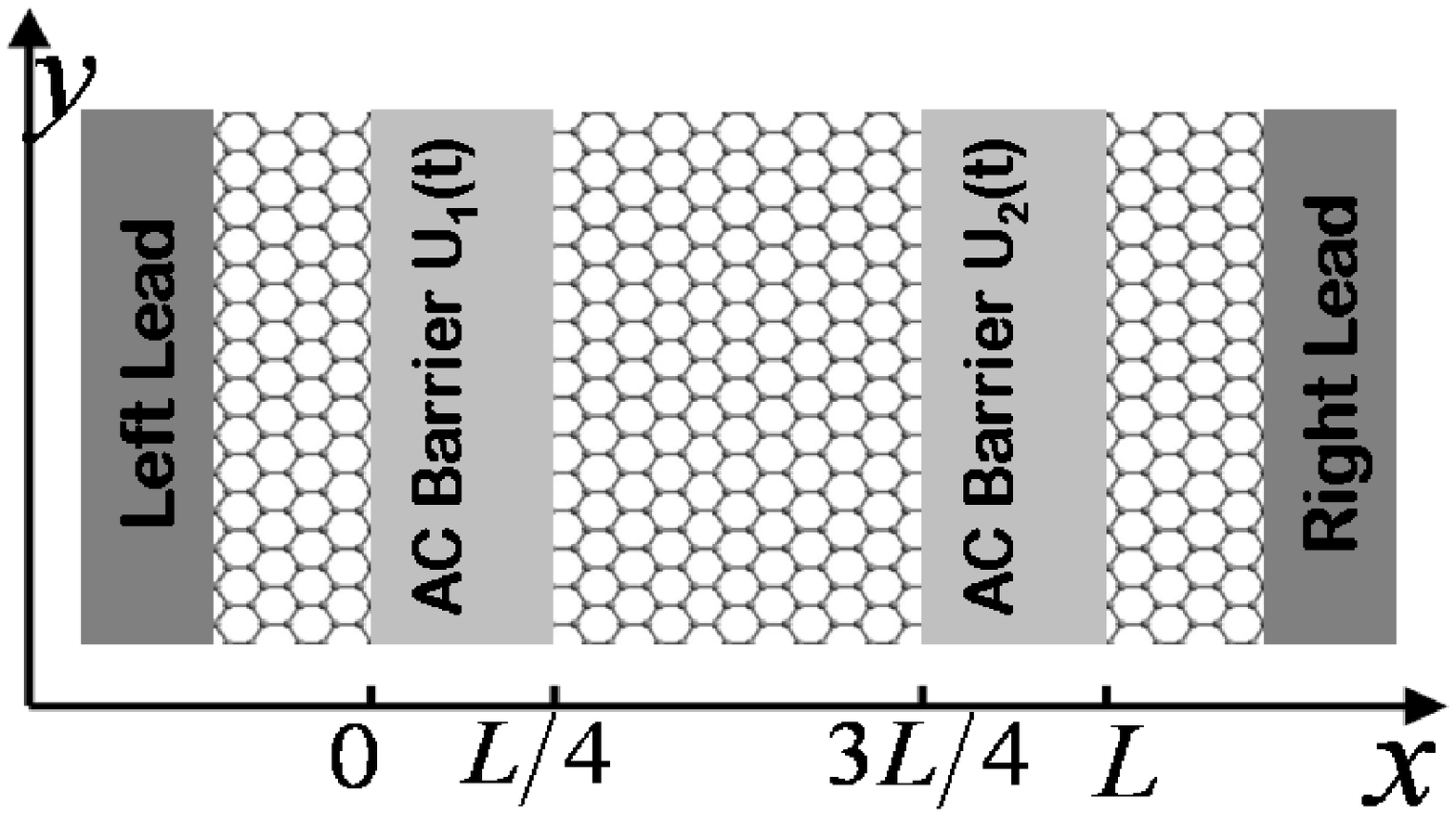}
\caption{Sketch of the quantum pump with ac-driving-force-modulated
double barriers in graphene.}
\end{figure}

\clearpage

\begin{figure}[h]
\includegraphics{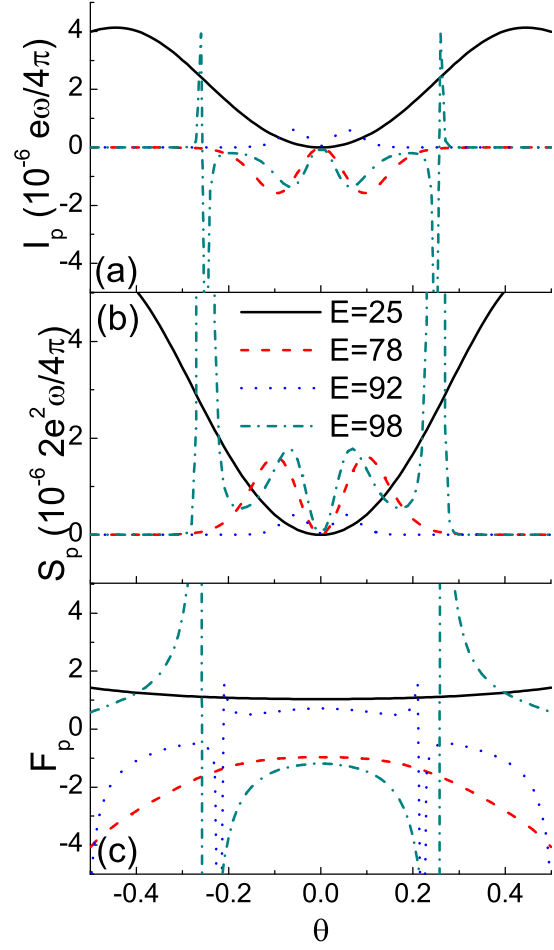}
\caption{Pumped current (a), shot noise (b), Fano factor (c) as
functions of the incident angle for different Fermi energies.
Driving amplitude $U_{\omega 1}=U_{\omega 2}=0.01$ meV. Driving
phase $\varphi _{1}=0.1$ and $\varphi _{2}=0.6$. The Fermi energy is
measured in meV.}
\end{figure}

\clearpage

\begin{figure}[h]
\includegraphics{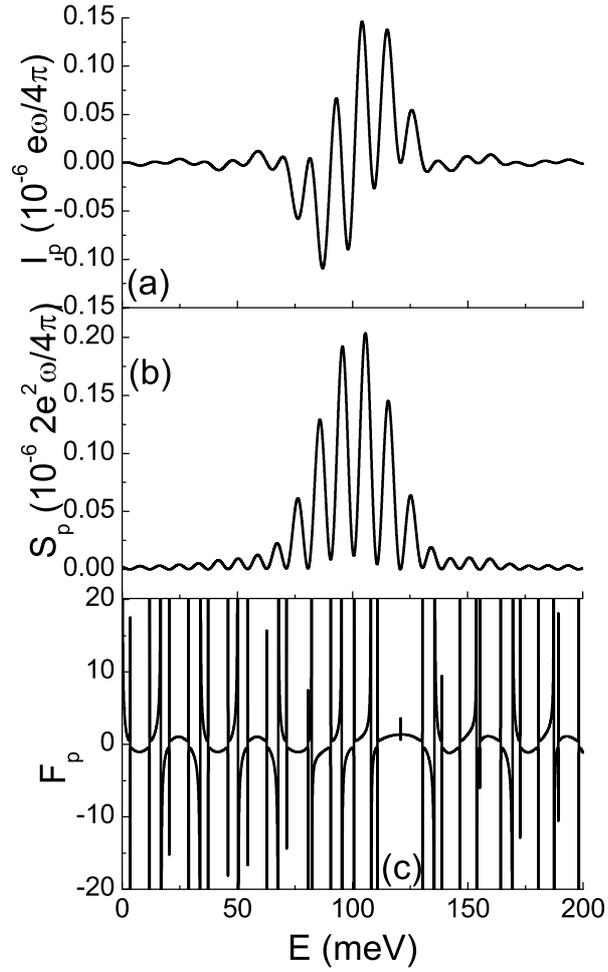}
 \caption{Pumped current (a), shot noise (b), Fano factor (c) as
functions of the Fermi energy. Driving amplitude $U_{\omega
1}=U_{\omega 2}=0.01$ meV. Driving phase $\varphi _{1}=0.1$ and
$\varphi _{2}=0.6$. Incident angle $\theta =0.01$.
 }
\end{figure}

\begin{figure}[h]
\includegraphics{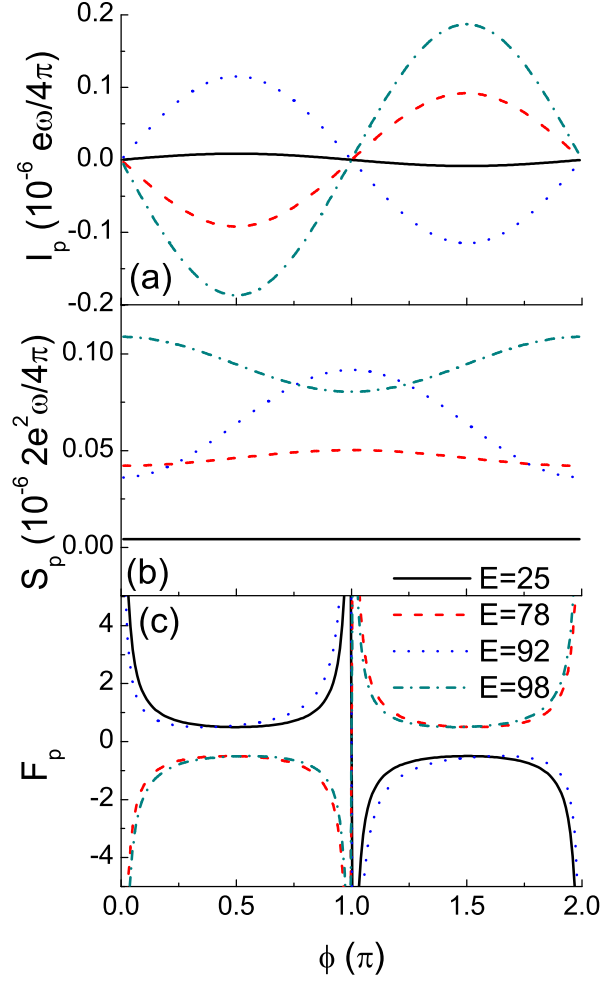}
 \caption{Pumped current (a), shot noise (b), Fano factor (c) as
functions of the driving phase difference for different Fermi
energies. Driving amplitude $U_{\omega 1}=U_{\omega 2}=0.01$ meV.
Incident angle $\theta =0.01$. The Fermi energy is measured in meV.}
\end{figure}

\end{document}